
\input harvmac \noblackbox

\def\p{^\prime}
\def\pp{^{\prime\prime}}
\def\d{\partial_{\pm}}


%
%
\def\RF#1#2{\if*#1\ref#1{#2.}\else#1\fi}
\def\NRF#1#2{\if*#1\nref#1{#2.}\fi}
\def\refdef#1#2#3{\def#1{*}\def#2{#3}}
\def\rdef#1#2#3#4#5{\refdef#1#2{#3, `#4', #5}}

%
%
\def\ts{\hskip .16667em\relax}

\def\CMP{{\it Comm.\ts Math.\ts Phys.\ts}}

\def\FAP{{\it Funct.\ts Analy.\ts Appl.\ts}}
\def\IJMP{{\it Int.\ts J.\ts Mod.\ts Phys.\ts}}

\def\JP{{\it J.\ts Phys.\ts}}

\def\NP{{\it Nucl.\ts Phys.\ts}}
\def\PL{{\it Phys.\ts Lett.\ts}}

\def\PR{{\it Phys.\ts Rev.\ts}}

\def\TMP{{\it Theor.\ts Math.\ts Phys.\ts}}

\def\Zm{Zamolodchikov}
\def\AZm{A.B. \Zm}

\def\ped{P.\ts E.\ts Dorey}
\def\dur{H.\ts W.\ts Braden, E.\ts Corrigan, \ped\ and R.\ts Sasaki}
%
%
\rdef\rAFZa\AFZa{A.E. Arinshtein, V.A. Fateev and \AZm}
{Quantum S-matrix of the 1+1 dimensional Toda chain}
{\PL {\bf B87} (1979) 389}

\rdef\rACFGZa\ACFGZa{H.\ts Aratyn, C.P.\ts Constantinidis, L.A.\ts Ferreira,
J.F.\ts Gomes and A.H.\ts Zimerman}
{Hirota's solitons in affine and the conformal affine Toda models}
{\NP {\bf B406} (1993) 727}

\rdef\rBCDRa\BCDRa{P. Bowcock, E. Corrigan, \ped\ and R.H. Rietdijk}
{Classically integrable boundary conditions for affine Toda field theories}
{in preparation}

\refdef\rBCDSa\BCDSa{\dur, \PL {\bf B227} (1989) 411}

\rdef\rBCDSc\BCDSc{\dur}
{Affine Toda field theory and exact S-matrices}{\NP {\bf B338} (1990) 689}

\rdef\rBCDSe\BCDSe{\dur}
{Multiple poles and other features of affine Toda field theory}
{\NP {\bf B356} (1991) 469}

\rdef\rBSa\BSa{H.\ts W.\ts Braden and R.\ts Sasaki}
{The S-matrix coupling dependence for a, d and e affine Toda
field theory}
{\PL {\bf B255} (1991) 343}

\rdef\rCDSb\CDSb{E.\ts Corrigan, \ped\ and R.\ts Sasaki}
 {On a generalised bootstrap principle}
 {\NP {\bf B408} (1993) 579--599}

\rdef\rCDRSa\CDRSa{E.\ts Corrigan, \ped\ , R.H.\ts Rietdijk and R.\ts Sasaki}
{Affine Toda field theory on a half line}
{\PL {\bf B333} (1994) 83}

\rdef\rCo\Co{I.\ts V.\ts Cherednik}
{Factorizing particles on a half line and root systems}
{\TMP {\bf 61} (1984) 977}

\rdef\rCMa\CMa{P.\ts Christe and G.\ts Mussardo}
{Integrable systems away from criticality: the Toda field theory and S
matrix of the tricritical Ising model}
{\NP {\bf B330} (1990) 465}

\rdef\rCMb\CMb{P.\ts Christe and G.\ts Mussardo}
{Elastic S-matrices in (1+1) dimensions and Toda field theories}
{\IJMP {\bf A5} (1990) 4581}

\refdef\rCTa\CTa{S.\ts Coleman and H.\ts Thun, \CMP {\bf 61}
 (1978) 31}

\rdef\rDf\Df{\ped }
{Root systems and purely elastic S-matrices, I \& II}
{\NP {\bf B358} (1991) 654; \NP {\bf B374} (1992) 741}

\rdef\rDi\Di{\ped}
{A remark on the coupling-dependence in affine Toda field theories}
{\PL {\bf B312} (1993) 291}

\rdef\rDHNa\DHNa{R.F. Dashen, B. Hasslacher and A. Neveu}
{}
{\PR {\bf D10} (1974) 414; \PR {\bf D11} (1975) 3424; \PR {\bf D12} (1975)
2443}

\rdef\rDGZb\DGZb{G.\ts W.\ts Delius, M.\ts T.\ts Grisaru and D.\ts Zanon}
{Quantum conserved currents in affine Toda theories}
{\NP {\bf B385} (1992) 307}

\rdef\rDGZa\DGZa{G.\ts W.\ts Delius, M.\ts T.\ts Grisaru and D.\ts Zanon}
 {Exact S-matrices for non simply-laced affine Toda theories}
 {\NP {\bf B282} (1992) 365}

\rdef\rDMSa\DMSa{G. Delfino, G. Mussardo and P. Simonetti}
{Statistical models with a line of defect}
{\PL {\bf B328} (1994) 123}

\refdef\rDSa\DSa{V.\ts G.\ts Drinfel'd and V.\ts V.\ts Sokolov,
 {\it J. Sov. Math.} {\bf 30} (1984) 1975}

\rdef\rFKc\FKc{A.\ts Fring and R.\ts K\"oberle}
{Factorized scattering in the presence of reflecting boundaries}
{\NP {\bf B421} (1994) 159}

\rdef\rFKd\FKd{A.\ts Fring and R.\ts K\"oberle}
{Affine Toda field theory in the presence of reflecting boundaries}
{\NP {\bf B419} (1994) 647}

\rdef\rFKe\FKe{A.\ts Fring and R.\ts K\"oberle}
{Boundary bound states in affine Toda field theory}
{Swansea preprint SWAT-93-94-28; hep-th/9404188}

\rdef\rGd\Gd{S.\ts Ghoshal}
{Boundary state boundary $S$ matrix of the sine-Gordon model}
{Rutgers preprint RU-93-51; hep-th/9310188}

\rdef\rGZa\GZa{S.\ts Ghoshal and \AZm}
{Boundary $S$ matrix and boundary state in two-dimensional
integrable quantum
field theory}
{Rutgers preprint RU-93-20; hep-th/9306002; to appear in \IJMP\ A}

\rdef\rHa\Ha{T.J. Hollowood}
{Solitons in affine Toda field theories}
{\NP {\bf B384} (1992) 523}

\rdef\rHb\Hb{T.J. Hollowood}
{Quantum soliton mass corrections in $SL(n)$ affine Toda field theory}
{\PL {B300} (1993) 73}

\rdef\rLa\La{G.L. Lamb, Jr}
{Elements of Soliton Theory}
{John Wiley and Sons Inc. 1980}

\rdef\rMv\Mv{A. MacIntyre}
{Integrable boundary conditions for classical sine-Gordon theory}
{in preparation}

\rdef\rMOPa\MOPa{A. V. Mikhailov, M. A. Olshanetsky and A. M. Perelomov}
{Two-dimensional generalised Toda lattice}
{\CMP {\bf 79} (1981) 473}

\rdef\rNa\Na{M.\ts R.\ts Niedermaier}
{The quantum spectrum of conserved charges in affine Toda theories}
{\NP {\bf 424} (1994) 184}

\rdef\rOTb\OTb{D.\ts I.\ts Olive and N.\ts Turok}
{The Toda lattice field theory hierarchies and zero-curvature
conditions in Kac-Moody algebras}
{\NP {\bf B265} (1986) 469}

\rdef\rRa\Ra{R. Rajaraman}
{Solitons and Instantons}
{North Holland 1982}

\rdef\rSk\Sk{R.\ts Sasaki}
{Reflection bootstrap equations for Toda field theory}
{Kyoto preprint YITP/U-93-33; hep-th/9311027}

\rdef\rSl\Sl{E.\ts K.\ts Sklyanin}
{Boundary conditions for integrable equations}
{\FAP {\bf 21} (1987) 164}

\rdef\rSm\Sm{E.\ts K.\ts Sklyanin}
{Boundary conditions for integrable quantum systems}
{\JP {\bf A21} (1988) 2375}

\rdef\rSZa\SZa{R.\ts Sasaki and F.\ts P.\ts Zen}
{The affine Toda S-matrices vs perturbation theory}
{{\it Int. J. Mod. Phys.} {\bf 8} (1993) 115}

\rdef\rTa\Ta{V.\ts O.\ts Tarasov}
{The integrable initial-value problem on a semiline:  nonlinear
\hfill\break Schr\"odinger
and sine-Gordon equations}
{{\it Inverse Problems} {\bf 7} (1991) 435}

\rightline{DTP-94/29}
\rightline{hep-th/9407148}
\medskip
\centerline{\bf Aspects of affine Toda field theory on a half line}
\bigskip
\centerline{E. Corrigan, P.E. Dorey, R.H. Rietdijk}
\bigskip
\centerline{Department of Mathematical Sciences}
\centerline{University of Durham, Durham DH1 3LE, England}
\medskip

\vskip 1in
Invited talk at \lq Quantum field theory, integrable models and beyond',
Yukawa Institute for Theoretical Physics, Kyoto University,
14-18 February 1994.
\vskip .5in

\centerline{\bf Abstract}
\bigskip
The question of the integrability of real-coupling
affine toda field theory on a
half line is discussed. It is shown, by examining low-spin conserved
charges, that the boundary conditions preserving integrability are
strongly constrained. In particular, among the cases treated so far,
$e_6^{(1)}$, $d_n^{(1)}$ and $a_n^{(1)}, \ n\ge 2$,
there can be no free parameters introduced by such boundary conditions;
indeed the only remaining freedom (apart from choosing the simple
condition $\partial_1\phi =0$), resides in a choice of signs. For a
special case of the boundary condition, accessible only for $a_n^{(1)}$,
it is pointed out  that the
classical boundary bound state spectrum may be related to a
set of reflection factors in the quantum field theory. Some
preliminary calculations are reported for other boundary conditions,
demonstrating that the classical scattering data satisfies the
weak coupling limit of the reflection bootstrap equation.

\vfill
\noindent July 1994\eject

\newsec{Introduction}

Over the last few years, there has been some progress in understanding
the particle scattering  of real coupling affine Toda field
theory~\NRF\rAFZa\AFZa\NRF\rBCDSc{\BCDSc\semi\BCDSe}%
\NRF\rCMa{\CMa\semi\CMb}%
\NRF\rBSa{\BSa\semi\SZa}%
\NRF\rDf{\Df}%
\NRF\rDGZa{\DGZa\semi\CDSb\semi\Di}\refs{\rAFZa{-}\rDGZa}.
Until
recently, the theory has been formulated on the full line $|x^1|\le\infty$.
However, it might be interesting for several reasons to consider the
theory defined on a half line, or on an interval, in which case
it becomes important to consider carefully the effect of adding
boundary conditions at one, or two, points. Consider, for
example, the restriction to a half line. Affine Toda field theory
is a lagrangian field theory and the addition of a boundary
(at $x^1=0$) requires a lagrangian of the form
\eqn\ltodahalf{\bar{\cal L}=\theta(-x^1){\cal L}-\delta(x^1){\cal B},}
where ${\cal B}$, which
is taken to be a functional of the fields but not their derivatives,
represents the boundary condition. In other words, at the boundary $x^1=0$
\eqn\todaboundary{{\partial\phi\over\partial x^1}=-{\partial{\cal B}
\over \partial\phi}.} The first term in \ltodahalf\ contains
the usual affine Toda lagrangian~\NRF\rMOPa{\MOPa}\refs{\rMOPa}:
\eqn\ltoda{{\cal L}={1\over 2}
\partial_\mu\phi^a\partial^\mu\phi^a-V(\phi )}
where
\eqn\vtoda{V(\phi )={m^2\over
\beta^2}\sum_0^rn_ie^{\beta\alpha_i\cdot\phi}.}
In \vtoda , $m$ and $\beta$ are real constants,
$\alpha_i,\ i=1,\dots ,r$ are the simple roots of some
Lie algebra $g$,
and $\alpha_0=-\sum_1^rn_i\alpha_i$ is an integer
linear combination of the simple roots; it corresponds to
the extra spot
on an extended (untwisted or twisted) Dynkin-Kac diagram
for $\hat g$. The coefficient
$n_0$ is taken
to be one.

Affine Toda field theory on the full line is classically integrable
\NRF\rOTb{\OTb}\refs{\rMOPa ,\rOTb};
there is a Lax pair representation of the field equations and,
as a consequence of this, there are infinitely many conserved charges in
involution. The charges $Q_s$ are labelled by spins $s$ which take values
equal to the exponents of the Lie algebra $g$ modulo its Coxeter element.
On the half line, translation invariance is obviously lost and the best
one can hope for by way of conserved quantities is some modification of
the parity even combination $ Q_s+ Q_{-s}$. For example, the energy is
$\hat Q_1+ \hat Q_{-1}+{\cal B}$,
(hatted quantities are integrated densities on the half line),
and is conserved whatever the form of the boundary condition might be.

Some time ago, Sklyanin and others~\NRF\rSl{\Sl\semi\Sm\semi\Ta}\refs{\rSl},
and, more recently,
Ghoshal and
Zamolodchikov~\NRF\rGZa{\GZa}\NRF\rGd{\Gd}\refs{\rGZa ,\rGd}
considered the question of the sine-Gordon
theory with a
boundary of exactly this type. The conclusion finally arrived at seems
to be that the most general boundary condition permitted by classical
considerations~\NRF\rMv{\Mv}\refs{\rMv} has the form:
\eqn\sgboundary{{\partial\phi\over\partial x^1}={a\over\beta}\sin \beta
\left({\phi -\phi_0 \over 2}\right)\qquad \hbox{at}\qquad x^1=0,}
where $a$ and $\phi_0$ are arbitrary constants, and $\beta$
is the sine-Gordon coupling
constant. Clearly, to preserve integrability,
the form of the boundary term in the lagrangian is
strongly constrained.

In affine Toda field theory, it was shown
recently~\NRF\rCDRSa{\CDRSa}\refs{\rCDRSa},
for the case
$g=a_n$, that the condition at the boundary is generally
even more restricted than it is for the sine-Gordon situation; indeed,
there is only a discrete ambiguity in the choice of ${\cal B}$.
It was found that the general form of
the boundary condition must be
\eqn\allboundary{{\cal B}={m\over \beta^2}\sum_0^rA_ie^{{\beta\over 2}
\alpha_i\cdot\phi},}
where the coefficients $A_i,\ i=0,\dots ,r$ are a set of real numbers.
Moreover, for  $  n>1 $ there was found to be an extra
constraint on the coefficients, namely:

\eqn\anboundary{\hbox{\bf either}\ |A_i|=2, \ \hbox{for}\ i=0,\dots ,n\
\hbox{\bf or}\ A_i=0\ \hbox{for}\ i=0,\dots ,n\ .}

Subsequent investigation has led to a general statement expected to be
applicable to
all simply-laced affine Toda theories:

\eqn\gboundary{\hbox{\bf either}\ |A_i|=2\sqrt{n_i}, \ \hbox{for}
\ i=0,\dots ,r\
\hbox{\bf or}\ A_i=0\ \hbox{for}\ i=0,\dots ,r\ .}

The sine- and sinh-Gordon theories appear to be the only ones for
which there is
a continuum of possible boundary conditions; for the others, the only
ambiguity at the boundary (up to an additive constant) is a choice of signs.

The argument leading to the necessity of
\anboundary\ required only that spin 2 charges  be preserved
(in the sense that there
is a boundary term $\Sigma_2$ for which $\hat Q_2+\hat Q_{-2}-
\Sigma_2$ is conserved).
For \gboundary , charges of higher spin
must be invoked, while a complete argument requires a generalisation of
the Lax pair idea
taking  into account the existence of the boundary.

Notice that for every choice of Lie algebra other than $a_n$,
the boundary condition
\todaboundary , together with the form of ${\cal B}$, fails
to  permit the vacuum solution
$\phi =0$, unless ${\cal B}=0$. This fact alone indicates that in general
the quantum field theory will be considerably complicated by the addition of
the boundary term. However, it is also this fact which makes the theory with
a boundary an interesting challenge.

As a final remark, notice that the permissible boundary conditions
fall into several types in the following sense. The original affine Toda
lagrangian is
invariant under translations of the field by vectors of the lattice dual to
the root lattice of the Lie algebra:
\eqn\fieldtransf{\phi \rightarrow \phi +{2\pi i\over \beta}\lambda
\qquad \hbox{where}\
\lambda\cdot\alpha_i = \hbox{integer}.}
The effect of such a transformation on the field theory with a boundary
condition \allboundary\ is to relate different boundary conditions
by altering the relative signs of the terms in ${\cal B}$. Clearly, the
term depending on the extra affine root $\alpha_0$ has a sign which is
determined once the other sign changes have been made by using a
suitable choice
of the vector $\lambda$. For example, in the $a_n$ theory boundary
conditions which differ by an even number of sign changes are related by
a transformation of the type \fieldtransf ; in this sense, there are just two
types. While the transformation \fieldtransf\ does not preserve the reality
of the Toda fields, it does  relate different ground states in
the complex version of the theory and is important for understanding the
existence of the complex solitons~\NRF\rHa{\Ha}\refs{\rHa}.

\newsec{Factorisable scattering with a boundary}

{}From the point of view of the quantum field theory, the first question
to ask is
what might be the optimal scenario if the theory with a
boundary remains integrable in the quantum domain. It is believed that affine
Toda field theory on the full line, restricted to real coupling,
is a relatively simple field theory with a
spectrum of $r$ scalar particles, whose scattering is purely elastic and
factorisable. Conjectures have been made for their spectrum and S-matrices;
these are consistent with the bootstrap idea and compatible with low order
perturbation theory \refs{\rAFZa{-}\rBSa}.
In fact, the theories
split naturally into two classes: those based on the \lq self-dual'
affine Dynkin-Kac diagrams ($a^{(1)}, d^{(1)}, e^{(1)}$ and
$a_{\rm even}^{(2)}$), and the rest which fall into dual pairs \refs{\rDGZa}.
For example,
for the self-dual theories, the quantum spectrum of conserved charges
(masses etc), and the quantum couplings are essentially identical to
the corresponding classical data.

Following the old
ideas of Cherednik~\NRF\rCo{\Co}\refs{\rCo}, the best one could expect
in the theory
including a boundary is that the particle spectrum remains the same, and
that far from the boundary the particles scatter
as if the boundary were absent.
The extra ingredient on the half line is that a particle approaching
the boundary is scattered elastically from it, with its $in$ and $out$
states related by a reflection factor (and of course a reversal
of momentum). Ie, in an integrable theory with distinguishable
scalar particles it might be expected that
\eqn\rfactordef{|a,\ -\theta_a >_{\rm out}=K_a(\theta_a )|a,
\  \theta_a >_{\rm in},}
where $a$ labels the particle, and $\theta_a$ is its rapidity.
Moreover, if the scattering theory remains factorisable, then
the scattering of any set of particles from the boundary can be
computed using the set of one-particle reflection factors, and the set
of two-particle S-matrices. The addition of the
boundary may also influence the spectrum by introducing new
kinds of states.

It must be remembered that it is
one thing to say what might be expected in the best possible case, and
quite another to establish it.
It is quite possible that this scenario only pertains to a subset
of the classical boundary conditions, or indeed to none of them;
the question remains to be settled.

On the other hand, if it is supposed that the scattering with a boundary
is factorisable in the manner described, then it is possible to generate a
number of consistency relations, based on the bootstrap,
which ought to be satisfied by the reflection factors $K_a$. These have
been formulated and explored recently by Fring
and K\"oberle~\NRF\rFKc{\FKc\semi\FKd}%
\refs{\rFKc}, by Sasaki~\NRF\rSk{\Sk}\refs{\rSk},
and by Ghoshal and
Zamolodchikov~\refs{\rGZa}, the latter in a more general context. It has
proved possible to conjecture a variety of solutions to the \lq
reflection bootstrap equations', some of which are described
in these references.
In \refs{\rCDRSa} the question of finding solutions
of the affine Toda bootstrap related to actual boundary
conditions was addressed,
and some suggestions were made concerning the spectrum of
boundary bound states
which might help to identify the  reflection factors for a particular
boundary condition among the plethora of solutions which
have been found
to the reflection bootstrap.

The relevant consistency conditions to be satisfied by
the reflection factors are:

\eqn\crossing{K_a^0(\theta_a )K_{\bar a}^0(\theta_a -i\pi )=  S_{aa}(2\theta_a
),}
and
\eqn\bootstrap{K_c^0(\theta_c ) = K_a^0(\theta_a )   K_b^0(\theta_b )
S_{ab}(\theta_a +\theta_b ).}
The second relation corresponds to the bulk coupling $ab\rightarrow c$
which appears as a
forward channel bound state pole in the  scattering of
particles $a$ and $b$, at
a particular relative rapidity
(for which $\theta_a=\theta_c -i\bar\theta_{ac}^b, \
\theta_b =\theta_c +i\bar\theta_{bc}^a   $),  corresponding
to the total energy and momentum of $a$ and $b$
coinciding with that of particle $c$.
The usual conventions have been
adopted  ($\bar\theta =\pi -\theta$). In eqs\crossing ,\bootstrap\
the superscript $0$ refers to the ground state of the boundary.

When the bootstrap is applicable and, as here, the scattering
is diagonal,
the first relation \crossing\ follows from repeated
application of the second, eq\bootstrap .
Note, though, for the theory corresponding to $a_1$ there are no
relations of the second type.

It is interesting to remark that in a weak coupling limit,
the particles become free far from the boundary
and $S_{ab}\rightarrow 1$.
However, particles must rebound from the boundary even
if they are free far away from it; therefore the small coupling
limit of the reflection
factors need not be unity. Taking the limit as the coupling tends to zero for
eq\bootstrap\ reveals that the \lq classical' coupling factors ought to
satisfy a bootstrap of their own, namely
\eqn\classbootstrap{K(\theta _c)=K(\theta_a)K(\theta_b),}
where the rapidities are related to each other as for \bootstrap .
Examples will be given below
which demonstrate that this remark is not a trivial one. Indeed,
it corresponds to a phenomenon which may be of interest to
anyone who has studied inverse scattering theory.

\newsec{Higher spin charges}

The spin even charges in $a_n$ theories play an important r\^ole
because they discriminate between mass degenerate conjugate particles.
If they are not preserved on the half line, the boundary may allow
the particles to mix with their conjugates on the rebound.

In the absence of a full Lax pair treatment of the half line problem,
the  pedestrian approach  adopted in \refs{\rCDRSa} will be employed
to explain the necessity of the constraints
\allboundary\ and \gboundary .
A fuller (and more satisfying) treatment will be
described elsewhere~\NRF\rBCDRa{\BCDRa}\refs{\rBCDRa}.
For related discussions of the problem on the full line,
see for example~\NRF\rDGZb{\DGZb\semi\Na}\refs{\rDGZb}.

The spin $\pm 3$ densities corresponding to the spin $\pm 2$ charges
for the whole line may be described by the general formulae (using light-cone
coordinates $x^{\pm}=(x^0\pm x^1)/\surd 2$):
\eqn\spintwo{T_{\pm 3}={1\over 3}A_{abc}\d\phi_a\d\phi_b\d\phi_c+B_{ab}\d^2
\phi_a\d\phi_b,}
where the coefficients $A_{abc}$ are completely symmetric and the
coefficients
$B_{ab}$ are antisymmetric. For constructing conserved quantities, the
densities
must satisfy
\eqn\currenta{\partial_{\mp}T_{\pm 3}=\d \Theta_{\pm 1}}
and explicit calculation reveals
\eqn\currentb{\Theta_{\pm 1}=-{1\over 2} B_{ab}\d\phi_aV_b,
\qquad V_b={\partial V\over \partial\phi_b},}
with the constraint
\eqn\aba{A_{abc}V_a+B_{ab}V_{ac}+B_{ac}V_{ab}=0.}
Eq\aba\ implies
\eqn\abb{{1\over \beta} A_{ijk}+B_{ij}C_{ik}+B_{ik}C_{ij}=0,}
where it is useful to define
\eqn\abdef{A_{ijk}=A_{abc}(\alpha_i)_a(\alpha_j)_b(\alpha_k)_c,\qquad
B_{ij}=B_{ab}(\alpha_i)_a(\alpha_j)_b,}
and
$$C_{ij}=\alpha_i\cdot\alpha_j.$$
Eq\abb\ implies that $B_{ij}$ is very restricted: it is non-zero only for
the $a_n^{(1)}$ cases (as expected since the exponent 2 does not occur
elsewhere) and, in those cases, $B_{ij}=0$ except for
$j=i\pm 1\ {\rm mod} \ n+1$, and $B_{i-1\, i}=B_{i\, i+1},\ i=1,\dots , n+1$.

Rewriting the conditions \currenta\ in terms of the variables $x^0,x^1$,
\eqn\currentc{\partial_0 (T_{+3}-\Theta_{+1}\pm  (T_{-3}-\Theta_{-1})) =
\partial_1(T_{+3}+\Theta_{+1}\mp  (T_{-3}+\Theta_{-1})),}
implies that  the combination $(T_{+3}-\Theta_{+1} + T_{-3}-\Theta_{-1})$
is a candidate density for a conserved quantity on the half line if,
at $x^1=0$,
\eqn\tboundary{(T_{+3}+\Theta_{+1}-  T_{-3}-\Theta_{-1})=\partial_0\Sigma_2.}
Then, provided \tboundary\ is satisfied, the  charge $P_2$, given by
\eqn\pthree{P_2=\int_{-\infty}^0dx^1 (T_{+3}-\Theta_{+1} + T_{-3}-
\Theta_{-1})
-\Sigma_2}
is conserved.

Eq\tboundary\ is a surprisingly strong condition. Together with the
definitions
\spintwo\ and \currentb , it implies that $\Sigma_2$ does not exist
unless the following two conditions hold at $x^1=0$:
\eqn\conditiona{A_{abc}{\cal B}_a+2B_{ab}{\cal B}_{ac}
+2B_{ac}{\cal B}_{ab}=0,}
\eqn\conditionb{{1\over 3}A_{abc}{\cal B}_a{\cal B}_b{\cal B}_c+2B_{ab}
V_a{\cal B}_b=0.}
Both of these involve the boundary term. Comparing \conditiona\ with \aba\
reveals that the boundary term ${\cal B}$ must be equal to
$${m\over\beta^2}\sum_0^rA_ie^{{\beta\over 2}\alpha_i\cdot\phi },$$
apart from an arbitrary additive constant. The second condition,
eq\conditionb , is nonlinear in the boundary term and therefore provides
equations for the constant
coefficients $A_i$ in terms of the coefficients in the potential.
To analyse these equations, the term in $A_{abc}$ is best
eliminated using \abb , to yield:
\eqn\conditionc{{1\over 24}\sum_{ijk}(B_{ij}C_{ik}+B_{ik}C_{ij})A_iA_jA_k
e_ie_je_k=\sum_{ij}B_{ij}A_je_i^2e_j,}
where
$$e_i=e^{{\beta\over 2}\alpha_i\cdot \phi}.$$
Comparing the coefficients of the products of exponentials in \conditionc\
requires either $A_i=0$ for all $i$, or, $A_i^2=4$ for all $i$.

The spin two contribution from the boundary is
$$\Sigma_2= - \sqrt{2}B_{ab} \partial_0\phi_a {\cal B}_b,   $$
and $\hat Q_2 +\hat Q_{-2} -\Sigma_2$ is conserved.

For $a_n^{(1)}$, similar analysis of the spin three and four charges
does not lead to
any stronger constraints on the coefficients. In \refs{\rCDRSa}
it was stated that the spin three charges led to weaker
constraints, in the sense that they required the same
general form of the boundary condition \allboundary ,
but did not lead by themselves to \anboundary .  Unfortunately,
this conclusion was mistaken.
It has since come to light that, except for $a_1$, a
term had been missed. Once this is taken into account,
one is forced to the same conclusion as was reached in
the spin two case. For the theories based on $d_n^{(1)}$, spin
three is always a possibility; insisting that there be an adaptation of the
spin
three charge in the presence of a boundary condition requires the
coefficients in \allboundary\ to satisfy \gboundary . The spin
four charge in the $e_6^{(1)}$ theory similarly constrains
the boundary condition in that case.

While the existence of the first few conserved charges is not enough to
guarantee integrability, the fact that the rather different calculations for
spins two, three and four all lead to the same conditions \allboundary ,
\gboundary\ seems to be strong evidence for their sufficiency.

\newsec{Classical boundary bound states}

With the suggested boundary condition  \allboundary ,
the equations of
motion for the theory on a half line become
\eqn\equations{\eqalign{\partial^2\phi& =-{m^2\over \beta}\sum_0^r n_i\alpha_i
e^{\beta\alpha_i\cdot\phi}\qquad x^1<0\cr
\partial_1\phi &=-{m\over 2\beta}\sum_0^r A_i\alpha_ie^{{\beta\over 2}
\alpha_i\cdot\phi}\qquad \ \ \ x^1=0.\cr}}
With the conventions adopted above, the total conserved energy is given by
\eqn\energy{E=\int^0_{-\infty} {\cal E}dx\ +\ {\cal B},}
where ${\cal E}$ is the usual energy density for Toda field theory. The
competition between the two terms when ${\cal B}$ is negative
permits the existence of boundary bound states.

The coupling constant $\beta$ can be used to keep track of the scale of the
Toda field $\phi$, in which case it is appropriate to consider an expansion
of the field as a power series in $\beta$ of the following type:
\eqn\expansion{\phi = \sum_{-1}^\infty \beta^k \phi^{(k)}.}
Generally, the series starts at $k=-1$ since, with
the conventions adopted above, the leading term on the
right hand side of the boundary condition is of order $1/\beta$, and may be
non-zero. The first two terms of the series satisfy the equations:
\eqn\firstterm{\eqalign{\partial^2\phi^{(-1)}&=-{m^2}\sum_0^r n_i\alpha_i
e^{\alpha_i\cdot\phi^{(-1)}}\qquad x^1<0\cr
\partial_1\phi^{(-1)}&=-{m\over 2}\sum_0^r A_i\alpha_i
e^{{1\over 2}\alpha_i\cdot\phi^{(-1)}}\qquad x^1=0,\cr}}
and
\eqn\secondterm{\eqalign{\partial^2\phi^{(0)}&=-{m^2}\sum_0^r n_i\alpha_i
e^{\alpha_i\cdot\phi^{(-1)}}\alpha_i\cdot \phi^{(0)}\qquad x^1<0\cr
\partial_1\phi^{(0)}&=-{m\over 4}\sum_0^r A_i\alpha_i
e^{{1\over 2}\alpha_i\cdot\phi^{(-1)}}\alpha_i\cdot \phi^{(0)}\qquad x^1=0.
\cr}}
The linear equations for $\phi^{(0)}$ represent the small coupling
limit once the background has been taken into account.
Exceptionally, $\phi^{(-1)}=0$ is a solution to
\firstterm\ when the coefficients $A_i$ are equal. In view of \gboundary ,
such a situation can arise only for $a_n^{(1)}$.
Otherwise, $\phi^{(0)}$ satisfies a linear equation in the background
provided by $\phi^{(-1)}$. Since $\phi^{(-1)}$ represents the \lq ground'
state, it is expected to be time-independent and of minimal energy.

In the $a_n^{(1)}$ case,
when the  coefficients are chosen to be $A_i=A,\ i=0,\dots ,r$,
and the ground state is
assumed to be $\phi^{(-1)}=0$, eqs\secondterm\ reduce to a diagonalisable
system whose solution in terms of eigenvectors $\rho_a$ of the mass$^2$ matrix
may be written as follows:
\eqn\casea{\phi^{(0)}=\sum_{a=1}^r\rho_a (R_ae^{-ip_ax^1}+I_ae^{ip_ax^1})
e^{-i\omega_ax^0}\ ,}
where
$$M^2\rho_a=m^2\sum_0^rn_i\alpha_i\otimes\alpha_i\rho_a=m^2_a\rho_a,\qquad
\omega_a^2-p_a^2=m_a^2,$$
and the reflection factor is given by\foot{The notation for the
reflection factor is chosen to
agree with some earlier references; unfortunately, it also disagrees
with others.}
\eqn\reflection{K_a=R_a/I_a={ip_a+Am_a^2/4m\over ip_a-Am_a^2/4m},
\qquad a=1,\dots ,r.}

If $A=0$, it is clear from \reflection\ that $K_a=1$ and there are no
exponentially decaying solutions to the linear system. On the other hand,
if $A\ne 0$ the reflection coefficients \reflection\ have poles at
$$p_a=-i{Am_a^2\over 4m},$$
for which
$$\omega_a^2=m_a^2(1-{A^2m_a^2\over 16m^2}).$$
Thus, provided $A^2<16m^2/m_a^2$ and $A<0$, the channel labelled $a$
has a bound state, with the corresponding
solution to the linear system decaying exponentially away from the
boundary as $x^1\rightarrow -\infty$.

For the case $a_n^{(1)}$, it has been established that $A^2=4$,
and the masses for the affine Toda theory are known to be
\eqn\anmasses{m_a=2m\sin({a\pi\over n+1}).}
Hence, with all the $A_i=-2$ , there are bound states for each $a$, with
\eqn\boundmasses{\omega^2_a=4m^2\sin^2({a\pi\over n+1})(1-
\sin^2({a\pi\over n+1}))=m^2\sin^2({2a\pi\over n+1}).}
Notice that there is a characteristic difference between $n$ odd
and $n$ even. In the latter case, the bound-state \lq masses' are
doubly degenerate, matching the degeneracy in the particle states
themselves. However, in the former case there is a four-fold degeneracy
in the bound-state masses, and $\omega_{(n+1)/2}=0$.

\newsec{Quantum boundary bound states}

One of the remarkable and intriguing
features of the quantum affine Toda field theories
based on  simply-laced Lie algebras is that the quantum mass spectrum
is essentially the same as the classical mass spectrum \refs{\rBCDSc ,\rCMa}.
It is
therefore tempting to suppose that a similar miracle might occur
for the theories on a half line, in which case the reflection factors
corresponding to the special boundary condition $A_i=-2$ (in the
case of $a_n^{(1)}$) will contain poles corresponding to the bound-state
masses \boundmasses . Since the S-matrices are known, the reflection
factors are strongly constrained (but not uniquely determined)
by the various bootstrap relations \bootstrap .

The simplest case to consider is $a_2^{(1)}$, which contains a conjugate
pair of particles with masses  given by
\eqn\atwodata{m_1=m_2=\sqrt{3}m.}
The classical reflection factors are given by \reflection , with $A=-2$.
It is useful to introduce the block notation (see \refs{\rBCDSc}
for details)
\eqn\block{(x)={\sinh({\theta\over 2}+{i\pi x\over 2h})\over
\sinh({\theta\over 2}-{i\pi x\over 2h})},}
where $h$ is the Coxeter number of the Lie algebra (in this case $h=3$).
In this notation,
the classical reflection factor is the same for
either particle and may be rewritten as follows:
\eqn\classfactor{{ip-{3m\over 2}\over ip+{3m\over 2}}=-(1)(2).}
Note that this expression does  satisfy the only classical
bootstrap relation \classbootstrap\ of the $a_2$ theory, corresponding
to the coupling $11\rightarrow 2$,  with $\theta_{11}^2
=2i\pi /3$.

In the same notation, the S-matrix elements are given by
\eqn\smatrix{S_{11}(\theta )=S_{22}(\theta )={(2)\over (B)(2-B)};\quad
  S_{12}(\theta ) =S_{11}(i\pi -\theta ) =-{(1)\over (1+B)(3-B)},}
  where the parameter $B$ depends on the coupling constant; it has been
  conjectured to have the form
  $$B(\beta )={\beta^2/2\pi \over 1+\beta^2/4\pi},$$
and checked to one-loop order for all simply-laced affine Toda theories
\refs{\rBSa}.
The boundary condition does not distinguish the two particles and, if
the two reflection factors describing reflection of either particle
off the ground state of the boundary are denoted $K_1^0(\theta )$ and
$K_2^0(\theta )$, it is expected that
$$K_1^0(\theta )= K_2^0(\theta ) .$$
In addition, the bootstrap equation \refs{\rFKc, \rGZa} consistent
with the $11\rightarrow 2$ coupling in the theory
\eqn\atwobootstrap{K_2^0(\theta ) = K_1^0(\theta -i\pi /3 )   K_1^0(\theta
+i\pi /3)  S_{11}(2\theta )}
must be satisfied, as must the \lq crossing-unitarity' relation~\refs{\rGZa}
\eqn\atwocrossing{K_1^0(\theta )K_2^0(\theta -i\pi )=  S_{11}(2\theta ).}

For the reason mentioned previously,  in this case
it is  sufficient to check the
bootstrap properties only; \atwocrossing\ is a consequence of the
bootstrap property.

On the basis of the discussion in the previous section, the reflection
factors are expected to contain a fixed simple pole (at $\theta =i\pi
/3$) indicating the
existence of the boundary bound state expected in each channel at the
mass $\sqrt{3}m/2$. It is also expected that as $\beta \rightarrow 0$
the
reflection factors revert to the classical expression \classfactor .
A \lq minimal' hypothesis with these properties is:
\eqn\quantfactor{K_1^0(\theta )= K_2^0(\theta )  = -{(1)(2+{B\over 2})
\over ({B\over 2})}.}
 Remarkably, this simple ansatz satisfies
both the requirements, \crossing\ and \bootstrap , as is easily
verified. As $\beta\rightarrow 0$,  the $\beta$-dependent factors
in \quantfactor\  give the
rapidity dependent factor $(2)$ in the classical reflection factor
\classfactor . This expression is not invariant under the transformation
$\beta\rightarrow
4\pi /\beta$, the weak-strong coupling symmetry characteristic of
the quantum affine Toda theory on the whole line.
Rather, as $\beta\rightarrow\infty$,
$K_1^0\rightarrow 1$.

Each channel has a boundary bound state (associated with the pole at
$\theta =i\pi /3$ ),
and it is convenient to label these
$b_1$ and $b_2$. The boundary bootstrap equation \refs{\rGZa}
defines the
reflection factors for the particles reflecting from the
boundary bound states.
If, as is being assumed here, there remain sufficiently
many charges conserved in the presence of the boundary to
ensure that the reflection off the boundary is diagonal, then
the equation given by Ghoshal and Zamolodchikov simplifies
dramatically. If the scattering of particle $a$ with the boundary
state $\alpha$
has a boundary bound state pole at $\theta =iv_{a\alpha}^\beta$,
then the reflection
factors for the new boundary state are given by
\eqn\newfactors{K^{\beta }_b(\theta )=S_{ab}(\theta -iv_{a\alpha}^\beta)
S_{ab}(\theta +iv_{a\alpha}^\beta)K^{\alpha }_b(\theta ).}
Thus, for the case in hand, the four possibilities are
\eqn\areflect{\eqalign {&K_1^{b_{1}}=S_{11}(\theta +i\pi /3)
S_{11}(\theta -i\pi /3)K_1^0(\theta )=S_{12}(\theta )K_1^0\cr
&K_2^{b_{1}}=S_{12}(\theta +i\pi /3)
S_{12}(\theta -i\pi /3)K_2^0(\theta )=S_{11}(\theta )K_2^0\cr}}
and
\eqn\breflect{\eqalign {&K_1^{b_{2}}=S_{12}(\theta +i\pi /3)
S_{12}(\theta -i\pi /3)K_1^0(\theta )=S_{11}(\theta )K_1^0\cr
&K_2^{b_{2}}=S_{11}(\theta +i\pi /3)
S_{11}(\theta -i\pi /3)K_2^0(\theta )=S_{12}(\theta )K_2^0.\cr}}
Consider the fixed pole structure of eqs\areflect .
Since both $S_{12}$ and $K_1^0$
have a simple pole at $\theta = i\pi /3$, their product
has a double pole;
this is not to be
interpreted as a new bound state. On the other hand, $S_{11}$ has a simple
pole
at $\theta = 2i\pi /3$ and $K_2^0$ has a simple pole at $\theta = i\pi /3$;
the first of these does not indicate a new boundary bound state since for
that interpretation $\theta$ ought to lie in the range $0\le\theta\le i\pi /2$.
However, the second pole lies in the correct range and indicates a boundary
state of mass $\sqrt{3}m$. This state has all the quantum numbers of particle
$1$ (the state $b_1$ has the quantum numbers of particle $2$ each
multiplied by $1/2$),
and may therefore be interpreted as a particle $1$ state at
zero momentum, next to the boundary in its ground state.
Establishing the latter
relies on the fact that the particle charges and the boundary state charges
are related in the quantum field theory via
\eqn\bbootstrap{P_s^a \cos (sv_{a\alpha}^\beta )= P_s^\beta -P_s^\alpha .}
 Eqs\breflect\
have a similar
interpretation. Consequently, it is tempting to conjecture
that the complete boundary spectrum corresponding to
the symmetrical boundary condition \allboundary\ with $A_1=A_2=-2$ consists
of a ground state, a pair of boundary states, and a tower of states
constructed by gluing zero rapidity particles to either the ground
state or to the boundary states $b_1,\ b_2$. Fring and K\"oberle
\NRF\rFKe{\FKe}\refs{\rFKe}
have reached  a similar pattern of boundary states by examining
a particular solution to the $e_6$ reflection bootstrap equations. They
do not, however, link their conjecture to a particular boundary condition.

On the other hand, if $A_1=A_2=2$, the classical reflection data has no
boundary bound states and the classical reflection coefficient \classfactor\
is replaced by its inverse. In this case, a candidate for the reflection
factors in the quantum field theory is
\eqn\quantfactora{K_1^0(\theta )= K_2^0(\theta )  = {(3-{B\over 2})
\over (2)(1-{B\over 2})}.}
This clearly satisfies all the bootstrap conditions and there are no
physical strip poles corresponding to boundary bound states. As $\beta
\rightarrow\infty$, these reflection factors tend to unity.

In order to generalise \quantfactor\ to other members of the $a_n$
series, it is useful to have some new notation. It is convenient to
introduce a pair of new blocks:
\eqn\newblock{<x>={(x+{1\over 2})\over (x-{1\over 2}+{B\over 2})},\quad
\widetilde{<x>}={(x-{1\over 2})\over (x+{1\over 2}-{B\over 2})}.}
These are related to the notation $[x]$ introduced in \refs{\rSk}
via
\eqn\ryublock{[x]=<x>\widetilde{<x>}.}
In terms of \ryublock , the quantities $S(2\theta )$ can be
manipulated conveniently, since
\eqn\stwotheta{\{ x\} (2\theta )=[x/2](\theta )/[h-x/2](\theta ),}
where
$$\{ x\} ={(x-1)(x+1)\over (x-1+B) (x+1-B)}$$
is the basic building block from which all the S-matrices of simply-laced
affine Toda field theories are constructed \refs{\rBCDSc}.

In terms of the new blocks, eq\quantfactor\ may be rewritten as
$$K^0_1={<{1\over 2}>\over <{5\over 2}>} ={<{1\over 2}>\over <h-
{1\over 2}>},$$
which is in a suitable form to generalise. Following the bootstrap,
using it  recursively to define all the other reflection factors,
leads to the general expression
\eqn\general{K_a^0={<a-{1\over 2}>\over <h-a+{1\over 2}>}
{<a-1-{1\over 2}>\over <h-a+1+{1\over 2}>}\cdots
{<{1\over 2}>\over <h-{1\over 2}>}=K_{h-a}^0.}
Moreover,
$$K_a^0\rightarrow -(a)(h-a),\qquad \beta\rightarrow 0$$
and, for each $a$, $K_a^0\rightarrow 1$ as $\beta\rightarrow\infty$.
The limit $\beta\rightarrow 0$ yields the classical reflection factor
\reflection , corresponding to particle $a$ in the field theory
based on $a_n^{(1)}$; this also satisfies \classbootstrap .
The generalisation of
\quantfactora\ is obtained by replacing $<x>$ by $\widetilde{<x>}$ in
\general .

This picture is quite attractive but tentative.
It must be emphasised that there is
much ambiguity involved in finding
solutions to the reflection bootstrap equation (see
\refs{\rSk}) and therefore, although it is tempting to use the
simplest expressions
satisfying the requirements, the solutions suggested
here are not guaranteed.
Even given a \lq minimality' assumption, it is quite
often the case that the solution is not unique. Further work needs to be done
to resolve these ambiguities.

For all other Toda theories, the classical background $\phi^{(-1)}=0$
is only possible for the relatively simple boundary condition ${\cal B}=0$.
This means that the above analysis is irrelevant in general and
either the only
integrable boundary condition for the other theories in the affine
Toda class is the
simple one, or one needs to develop a technology to explore
the non-trivial static backgrounds. These provide an effective potential
for the linearised scattering problem, corresponding to the weak
coupling limit, and the derived
scattering data provides a selection principle among the known solutions to
the bootstrap equations \bootstrap . Basically, these background solutions are
similar to solitons in the complex theory,
continued to real coupling. Such solutions are real and  inevitably
diverge at some value of $x^1$. Provided the singularity occurs
beyond the boundary, in $x^1>0$,
there  is no cause for alarm.

Obtaining the background field configurations
for a given boundary condition is a formidable problem and is certainly
unsolved for the general case. A few illustrative examples will be
given in the following sections.

\newsec{The $a_2^{(1)}$ theory with an asymmetrical
boundary condition}

As a first example, consider the case of
$a_2^{(1)}$  with an asymmetrical
boundary condition. For definiteness, take $A_1=2,\ A_2=A_0
=-2$.
For this case,
$\phi^{(-1)}=0$ is not an option and the first task
is to solve for the background configuration. The ansatz
\eqn\atwoansatz{\phi^{(-1)}=  \alpha_1 \rho}
is compatible with the boundary condition and leads to
a time-independent\foot{From now on $x\equiv x^1$} Bullough-Dodd equation
($a_2^{(2)}$ affine Toda):
\eqn\bd{\eqalign{&\rho\pp =e^{2\rho} - e^{-\rho}\qquad x<0\cr
                 &\rho\p =-(e^{\rho}+e^{-\rho /2})\qquad x=0.\cr}}
At first sight, \bd\ does not look promising, particularly the boundary
condition. The corresponding equations for $\phi^{(0)}$ are (putting $m=1$):
\eqn\phizero{\eqalign{\partial^2\phi^{(0)}&=-\sum_0^r n_i\alpha_i
e^{\alpha_i\cdot\phi^{(-1)}}\alpha_i\cdot \phi^{(0)}\qquad x^1<0\cr
&=-\pmatrix{ 2e^{2\rho}+e^{-\rho} &0\cr
              0& 3e^{-\rho}\cr}\phi^{(0)}\cr
\partial_1\phi^{(0)}&={1\over 4}\sum_0^r A_i\alpha_i
e^{{1\over 2}\alpha_i\cdot\phi^{(-1)}}\alpha_i\cdot \phi^{(0)}
\qquad x^1=0\cr
&=-{1\over 2}\pmatrix{2e^{\rho}-e^{-\rho /2} &0\cr
                               0 &-3 e^{-\rho /2}\cr}\phi^{(0)}.\cr}}
In the following analysis, it is convenient first to concentrate on
the second component of $\phi^{(0)}$, returning to the other later.

Integrating  the Bullough-Dodd equation once yields
$$(\rho\p)^2=e^{2\rho}+2e^{-\rho}-3,$$
and therefore at the boundary $x=0$
$$(e^\rho+e^{-\rho /2})^2=e^{2\rho}+2e^{-\rho}-3.$$
The latter implies
$$e^{\rho /2}=-1,\ 1/2, \ \hbox{or}\ \rho\rightarrow\infty .$$
Clearly, the first of these possibilities (which is a double root)
is incompatible with the reality of $\rho$ but the second of
them is fine; the third
implies a singularity at the origin which may be acceptable. In
fact, for the above choice of coefficients, the finite solution
is the appropriate one. Given this
boundary value the boundary condition for the linear approximation
around the background is:
\eqn\bdzero{\partial_1\phi^{(0)}=\pmatrix{3/4 &0\cr
                               0 &3\cr}\phi^{(0)}.}

The relevant solution to Bullough-Dodd, with $\rho\rightarrow 0$ as
$x\rightarrow -\infty$, is given by:
\eqn\bdsolution{e^{-\rho} ={1+4E+E^2\over 1-2E+E^2}=1+{3/2\over \sinh^2\sqrt{3}
(x-x_0)/2 }, \qquad E=e^{\sqrt{3}(x-x_0)},}
where $x_0$ is to be determined by the boundary condition. It must, however,
turn out that $x_0>0$ so that the singularity in \bdsolution\ does not
occur inside the half line $x<0$. To check this the boundary condition
is matched by
$$e^{-\rho}=4=1+(3/2)({\rm coth}^2 \sqrt{3}(x_0)/2\ -1),$$
ie
$${\rm coth}^2 \sqrt{3}(x_0)/2 =3.$$
The positive solution to this must be chosen, given the sign of
the slope of $\rho$ near the boundary. Note that the solution \bdsolution\
is a static solution of the type discovered by Aratyn et al
\NRF\rACFGZa{\ACFGZa}\refs{\rACFGZa}. Regarded as
a solution to the $a_2^{(1)}$ classical Toda equations, it is of  the
standard Hirota type but not of the kind originally discussed by Hollowood
\refs{\rHa}.

Next, consider the equation for $\phi^{(0)}$.
We expect to find a solution of the form
$$\phi^{(0)}=e^{-i\omega t}\Phi (x).$$
However, given the form of the background
potential, it is convenient to make a change of
variable to $z=\sqrt{3}x/2$. Once this has been done, $\Phi$ satisfies:
\eqn\Phiequation{\Phi\pp = \left( -\lambda^2 +
{6\over \sinh^2 (z-z_0)}\right)\Phi ,}
where it is convenient to set
$$\lambda^2=(4/3)(\omega^2 -3)=(4/3)p^2=4\sinh^2(\theta).$$
This is quite a striking result:  not only
is the \lq potential' on the right
hand side of the equation  in the class of exactly solvable ones, it has a
coefficient which, in the case of a $1/\cosh^2$ potential, indicates no
reflection. Therefore, the equation for $\Phi$ is very special, and
indeed solvable in elementary terms. The known solution (see
\NRF\rLa{\La}\refs{\rLa}, for example) has the form
\eqn\lamb{\Phi_L =\left( {d\over dz}-2\coth (z-z_0)\right)
\left( {d\over dz}-\coth (z-z_0)\right)e^{i\lambda z}}
leading to a general solution for $\Phi$ of the form
$$\Phi =a\Phi_L+\hbox{complex\ conjugate},$$
from which the reflection coefficient can be found. Firstly,  as $z\rightarrow
-\infty$, $\coth (z-z_0)\rightarrow -1$ and so:
\eqn\asymptotic{\Phi\sim a(i\lambda +2)(i\lambda +1)e^{i\lambda z}+
a^*(-i\lambda +2)(-i\lambda +1)e^{-i\lambda z}.}
Therefore, the reflection coefficient is given by
\eqn\rfactor{K={a^*\over a}{(i\lambda -2)\over (i\lambda +2)} {(i\lambda
-1)\over
(i\lambda +1)}.}
Next, \lamb\  and its derivative at $x=0$ need to be evaluated in
order to fix the relationship between $a$ and $a^*$. The
result is:
\eqn\zerovalues{\eqalign{\Phi_L(0)&=(i\lambda )^2 +3\sqrt{3} (i\lambda ) +8\cr
\Phi_L\p(0)&= (i\lambda )^3 +3\sqrt{3}(i\lambda )^2 +14 (i\lambda ) +12
\sqrt{3}.\cr}}
Therefore the boundary condition becomes:
$$a((i\lambda )^3+\sqrt{3}(i\lambda )^2-4(i\lambda )-4\sqrt{3})+cc =
a(i\lambda +\sqrt{3})(\, (i\lambda^2)-4)+ cc =0,$$
from which
$${a^*\over a}={i\lambda +\sqrt{3}\over i\lambda -\sqrt{3}}.$$
Remembering that $\lambda =2\sinh\theta \equiv 2s$ leads to
$$K={is+\sqrt{3}/2 \over is-\sqrt{3}/2}\ {is-1\over is+1}\ {is-1/2\over
is+1/2},$$
which, in the usual notation \block , is just
\eqn\rmixed{K_1={(1/2)(3/2)^2(5/2)\over (1)(2)(3)}.}

Remarkably, this reflection data corresponding to one of the channels
implies the same data in the other channel, assuming the \lq classical
reflection' bootstrap equation \classbootstrap\ holds.   Indeed, the
denominator is identical to data obtained above from the symmetrical boundary
condition, while the numerator satisfies the bootstrap on its
own. Clearly \rmixed\ can be regarded as the classical limit of
a solution to the full bootstrap equation in many ways. The simplest
such solution would be
\eqn\rmin{K_1=K_2=(1/2)(3/2)^2(5/2)\, {(3-{B\over 2})\over (1-{B\over 2})(2)},}
but there are many others. For example, if $C(\beta )$ is any function of
$\beta$, vanishing at $\beta =0$, then
$$(1/2 +C)(3/2-C)(3/2+C)(5/2-C) \, {(3-{B\over 2})\over (1-{B\over 2})(2)},$$
is also a solution.

A check on all of this is to calculate directly the reflection data in the
other channel and verify that it is indeed the same.
The linear approximation in the
background potential in the other channel  has the form
$$\Phi\pp =\left(-\lambda^2 +{4r\over q^2}\right)\Phi ,$$
where
$$\eqalign{r&= -6E(1-6E+3E^2+4E^3+3E^4-6E^5+E^6)\cr
           q&= 1+2E-6E^2+2E^3+E^4,\cr}$$
and, as before,
$$E=e^{\sqrt{3}(x-x_0)}=e^{2(z-z_0)}.$$
The equation is solved by  taking $\Phi$ to have the form
$$\Phi_L = {p\over q}\ e^{i\lambda z},$$
where ($i\lambda =2 i\sinh\theta$) and the function $p$ depends on $\lambda$
and is given up to an overall constant by
$$p=(2+i\lambda )(1+i\lambda )-2(\lambda^2 +4)(E+E^3)+6(2+\lambda^2)E^2+
(2-i\lambda )(1-i\lambda )E^4.$$
At $z=0$ the following boundary condition holds:
$$\Phi\p (0)={\sqrt{3}\over 2}\Phi (0).$$
Hence, setting
\eqn\other{\Phi =a{p\over q}\ e^{i\lambda z}+cc}
and imposing the boundary condition fixes the ratio $a^*/a$:
$${a*\over a}={i\lambda +\sqrt{3} \over i\lambda -\sqrt{3}}.$$
Using the latter, the reflection factor determined by the solution
\other\  is
$$K={i\lambda +\sqrt{3} \over i\lambda -\sqrt{3}}~{(2-i\lambda )(1-i\lambda )
\over (2+i\lambda )(1+i\lambda )},$$
which is precisely the same as \rmixed .

It is striking, and quite surprising, to find that the classical
reflection factors
obtained by the direct calculations described here do satisfy the
simple classical
bootstrap equation \classbootstrap ; that this is the case
provides some support
for the approach advocated above. The phenomenon is not confined to the
$a_n^{(1)}$ theories, as the next section will show.

\newsec{The $d_5^{(1)}$ theory}

As a second example, consider the $d_5^{(1)}$ theory
and solve the background equation in a symmetrical situation
(ie $A_0=A_1=A_4=A_5=-2, \ A_2=A_3=-2\sqrt{2}$). This choice is expected to
contain bound states; again,
$\phi^{(-1)}=0$ is not an option. Hence, there is no choice but to find the
background first.
There is still a symmetry in the sense of preserving the
symmetry of the affine Dynkin diagram for $d_5^{(1)}$, and the
ansatz\foot{The centre spots of the diagram are labelled \lq 2,3'.}
\eqn\dfiveansatz{\phi^{(-1)}=  (\alpha_2+\alpha_3) \rho}
is compatible with the boundary condition and
leads to the sinh-Gordon equation for $\rho$.
\eqn\bda{\eqalign{&\rho\pp =2(e^\rho - e^{-\rho})\qquad x<0\cr
                 &\rho\p = -2e^{-\rho /2} +\sqrt{2}e^{\rho /2}\qquad x=0.\cr}}
This looks straightforward but there is a subtlety.

Integrating the first of eqs\bda , and matching with  the second
at the boundary gives:
$$2e^{-\rho /2} -\sqrt{2}e^{\rho /2}=2(e^{-\rho /2}-e^{\rho /2}),$$
which implies that $e^{\rho /2}$ vanishes at the boundary. At first
sight, this  seems unreasonable. However, a calculation of the
energy of this field configuration gives zero; the half line
contribution and the boundary term (both infinite)  cancel precisely.
The relevant
solution to the sinh-Gordon equation is
\eqn\sg{e^{\rho /2}={1-e^{2x}\over 1+e^{2x}}.}

The equations for $\phi_0$ can now be computed and, in a suitable
basis for the roots, the field equation  becomes
\eqn\phizeroa{\partial^2\phi^{(0)} =-\left[
e^{-\rho} \pmatrix{1 & 0& 1&0 &0\cr 0& 2& 0&0&0\cr
1& 0& 1 &0&0\cr 0&0 & 0&2&0\cr
0&0&0&0&2\cr}+e^\rho\pmatrix{(1{-}\sqrt{2})^2&0&-1&0&0\cr
0&2&0&0&0\cr -1&0&(1{+}\sqrt{2})^2&0&0\cr
0&0&0&0&0\cr
0&0&0&0&0\cr}\right]\phi^{(0)}.}
Clearly, the only easily solvable equations are those corresponding
to the $s$, $\bar s$ particles of mass $\sqrt{2}$, and the particle $2$
of mass $2$. Since the
background potential is singular at $x=0$, an acceptable
solution must vanish at the origin. Using the expression
for the background field \sg , the linearised equation for the $s$ or
$\bar s$ components takes the form:
$$\partial^2\phi_s=-\left(2+{2\over\sinh^2x}\right)\phi_s,$$
which is again exactly solvable. Setting
$$\phi_s=e^{-i\omega t}\Phi_s,$$
the solution is given by
$$\Phi_s=a(i\lambda -\coth x)e^{i\lambda x} +cc \qquad \hbox{with}\
\lambda^2=\omega^2-2=2\sinh^2\theta .$$
Near the origin
$$\coth x={1\over x}(1+x^2/3+O(x^4));$$
therefore, the choice $a^*=-a$ is enough to remove the singularity. Indeed,
near the origin $\Phi_s=O(x^2)$ which is exactly right for ensuring
the boundary condition is also satisfied at $x=0$. In other words, the
singular behaviour of the background is harmless. Hence,
$$\phi_s= ae^{-i\omega t}\left[(i\lambda -\coth x)e^{i\lambda x} -cc\right] ,$$
from which the reflection data for the classical scattering may be read off,
to give
\eqn\sdata{K_s={i\lambda -1\over i\lambda +1}=-(2)(6)=K_{\bar s}.}
Again, the notation \block\ is convenient (with $h=8$).
If \sdata\ is taken seriously, it implies the presence of a
bound state at $\theta =i\pi /4$ with mass $m_s /\sqrt{2}$.

The other diagonal channel has an effective potential
$$-2(e^\rho +e^{-\rho})=-2(2+{4\over \sinh^2 2x}),$$
which is again integrable of the same type apart from the scale of $x$.
Calculating the classical scattering data yields
$$K_2={i\lambda -2\over i\lambda +2}=-(4)^2.$$
No bound state would be expected in this channel.

On the other hand, it can be
assumed that \sdata\ provides a starting point for a classical
bootstrap calculation; then the bootstrap itself will determine the scattering
data for the other particles, provided it is consistent. The $d_5$
couplings are:
$$ss1\ \ ss3\ \ s\bar s2\ \ 112\ \ 123\ \ 332$$
with the appropriate coupling angles given in \refs{\rBCDSc}.
Remarkably, \sdata\
provides a consistent solution to the classical bootstrap \classbootstrap\
based on these couplings for which
\eqn\dfivedata{\eqalign{K_s=K_{\bar s}&=-(2)(6)\cr
                     K_1&={(3)(5)\over (1)(7)}\cr
                     K_2&=-(4)^2\cr
                     K_3&=(1)(3)(5)(7)\cr}}
is the full set of data. Note that the expression calculated above for $K_2$
is consistent with the bootstrap. A lengthier calculation has been done
to check the reflection data in the channels corresponding to particles
$1$ and $3$. This will not be described here for reasons of space,
but makes use of the idea that
the linearised approximation may be thought of as a particular limit
of what would be a double soliton solution in the imaginary coupling theory
\NRF\rDHNa{\DHNa\semi\Ra}\NRF\rHb{\Hb}\refs{\rDHNa ,\rHb}.
At first sight it seems unlikely that the scattering data will be
diagonal, given the mixing in the linearised equation \phizeroa . However,
not only is the scattering diagonal, the computed reflection data
precisely matches the prediction of \dfivedata !

There are several possible solutions to the reflection
bootstrap equations for which \dfivedata\  are the \lq classical' limit.

\newsec{The $a_1^{(1)}$ or sinh-Gordon model}

As mentioned in section two, this case, in the sense of sine-Gordon,
has been studied before. It turns out that in the present context  a
reasonably complete analysis
may be made.

The equation for the static background is (for convenience
$\rho =\phi^{(-1)}$)
\eqn\sg{\eqalign{\rho^{\prime\prime}&=-\sqrt{2}\left(e^{\sqrt{2}\rho}-
e^{-\sqrt{2}\rho}\right)\qquad x<0\cr
\rho^\prime&=-\sqrt{2}\left(\epsilon_1e^{\rho /\sqrt{2}}-\epsilon_0e^{-\rho
/\sqrt{2}}\right)\qquad x=0, \qquad A_i=2\epsilon_i\cr}}
from which, on integrating the first equation once, and comparing with
the boundary condition, one obtains
\eqn\sga{\eqalign{\rho^\prime& =\sqrt{2}\left(e^{\rho /\sqrt{2}}-
e^{-\rho /\sqrt{2}}\right)\qquad x<0\cr
e^{\sqrt{2}\rho}&={1+\epsilon_0\over 1+\epsilon_1}\qquad x=0.\cr}}
Hence, the ground state solution has the form
\eqn\sgbackground{e^{\rho /\sqrt{2}}={1+e^{2(x-x_0)}\over
1-e^{2(x-x_0)}},}
with
\eqn\sgbv{\hbox{coth}\, x_0=\sqrt{1+\epsilon_0\over 1+\epsilon_1}.}
The expression given in \sgbv\ assumes $\epsilon_0>\epsilon_1$;
if that is not the case, it is necessary to adjust the solution by
shifting $x_0$ by $i\pi /2$.

The linearised wave equation in this background has the form
\eqn\sglin{\eqalign{\partial^2\phi^{(0)}&=-4\left(1+{2\over
\sinh^2 2(x-x_0)}\right)\phi^{(0)}\qquad x<0\cr
\partial_1\phi^{(0)} &=-\left(\epsilon_0\tanh x_0 +\epsilon_1
\coth x_0\right)\phi^{(0)}\qquad x=0.\cr}}

The classical scattering data for this potential
is computable  in terms
of the parameters in the boundary term. It is convenient to set
$\phi^{(0)}=e^{-i\omega t}\Phi(z)$, in which case the solution to \sglin\
takes the form
$$\Phi (z)=a(i\lambda -\coth (z-z_0))e^{i\lambda z} +cc,\qquad
\lambda =\sinh\theta ,$$
where the ratio of coefficients $a^*/a$ can be computed from the boundary
condition.  The reflection coefficient may be read off and turns out to be
\eqn\sgcoeff{\eqalign{K&= {1-i\lambda \over 1+i\lambda}\, {(i\lambda)^2+
i\lambda\sqrt{1+\epsilon_0}\sqrt{1+\epsilon_1}+(\epsilon_0+\epsilon_1)/2
\over (i\lambda)^2-
i\lambda\sqrt{1+\epsilon_0}\sqrt{1+\epsilon_1}+(\epsilon_0+\epsilon_1)/2}\cr
&=-\,
(1)^2~\left[(1+a_0+a_1)(1-a_0+a_1)(1+a_0-a_1)(1-a_0-a_1)\right]^{-1},\cr}}
where in the last step it has been convenient to set
$$\epsilon_i=\cos a_i\pi , \qquad |a_i|\le 1,\qquad i=0,1.$$
To extend beyond the restriction on the $a_i$, it is necessary to continue
the formula \sgcoeff\ by making the substitution $a_i\rightarrow a_i+2$.

The classical result \sgcoeff\ is quite elegant and in fact remarkably
similar to the expression for the quantum reflection factor suggested
by Ghoshal~\refs{\rGd} in the case of the scattering of the lightest
sine-Gordon
breather state from the boundary\foot{To make the comparison, a suitable
\lq classical' limit of Ghoshal's
formula must be taken after analytic continuation in $\beta$.}.

As a final check on the stability of the background,
it is interesting to examine the energy as a functional of the  field
$\phi^{(-1)}\equiv\rho$. Including
the boundary contribution, the energy is given by:
\eqn\sgenergy{E=\int_{-\infty}^0dx~\left({(\rho^\prime )^2\over 2} +
(e^{\sqrt{2}\rho}+
e^{-\sqrt{2}\rho}-2)\right)+A_1e^{\rho_0/\sqrt{2}}+A_0e^{-\rho_0/\sqrt{2}}.}
Using the Bogomolny argument, this may be rewritten,  replacing the
integrand by
$${1\over 2}\left(\rho^\prime -\sqrt{2}(e^{\rho /\sqrt{2}}-e^{-\rho /\sqrt{2}})
\right)^2+\sqrt{2}\rho^\prime (e^{\rho /\sqrt{2}}-e^{-\rho /\sqrt{2}}),$$
to obtain
\eqn\sgbound{E\ge \, -4+(A_0+2)e^{-\rho_0/\sqrt{2}}+(A_1+2)
e^{\rho_0/\sqrt{2}}.}
{}From this it is clear that the energy is definitely bounded below
provided both $A_0$ and $A_1$ are at least $-2$.

\newsec{Comments}

This is work in progress and there remains much to do. The principal
question is: what, if any, boundary conditions are compatible with
integrability? From a classical point of view, there are  strong
constraints on the permitted boundary conditions for affine Toda
field theory. From the point of view of the quantum field theory, the
situation is less clear and the story is far from complete.
On the assumption that a form of integrability
survives, such that scattering and reflection from the boundary remain
elastic and factorisable, it is possible to make a variety of conjectures
for the reflection factors. It is not clear to what extent any of the
solutions listed in \refs{\rFKc ,\rSk ,\rFKe }, or \refs{\rCDRSa}, can
be said to follow from a particular boundary condition. There are some
conjectures but in the end a proper formulation of the theory, via
perturbation theory or otherwise, will be needed to decide the issues.
On the other hand, the search for the ground state
corresponding to a particular
choice of boundary condition, and the study of this
linearised classical
problem, are of
interest in themselves. A proper understanding of the
classical problem would appear to be a necessary
prerequisite to a formulation of perturbation theory
in anything other than a
situation with a trivial condition at the boundary.
The discussion of a system
with two boundaries would be interesting from the point of view of
finite size effects,
and possibly for string theory, particularly in view of the extra states
in the spectrum of the theory once boundaries are included. Finally, the
boundary conditions considered here have been assumed to be impenetrable.
However, there are other possibilities~\NRF\rDMSa{\DMSa}\refs{\rDMSa}
involving the inclusion of internal boundaries,
impurities or defects, which allow transmission as well as reflection;
these are interesting in their own right, but
have been deliberately excluded from the work
discussed here.

\bigskip

\noindent{\bf Acknowledgements}

We have benefited from discussions with Peter
Bowcock, Alistair MacIntyre and Ryu Sasaki for which we are very grateful.
One of us (EC) is also grateful to the organisers of
the meeting  for giving him the opportunity to present some of these ideas;
another (RHR) gratefully acknowledges support from the UK Engineering and
Physical
Sciences Research Council.  This investigation
was supported in part by a Human Capital and Mobility grant of the
European Union, contract number ERBCHRXCT920069.

\listrefs
\end